# SKA shakes hands with *Summit*

Ruonan Wang[a], Andreas Wicenec[b,*], Tao An[c,*]

[a] *Oak Ridge National Laboratory (ORNL), Oak Ridge, TN 37830, USA*
[b] *International Centre for Radio Astronomy Research (ICRAR), The University of Western Australia, Crawley, WA 6009, Australia*
[c] *Key Laboratory of Radio Astronomy, Shanghai Astronomical Observatory (SHAO), Chinese Academy of Sciences, Shanghai 200030, China*

*Corresponding authors. Email addresses*: andreas.wicenec@icrar.org (A. Wicenec), antao@shao.ac.cn (T. An).

Recently, a full-scale data processing workflow (Fig. 1) of the Square Kilometre Array (SKA) Phase 1 was successfully executed on the world's fastest supercomputer *Summit*, proving that scientists have the expertise, software tools and computing resources to process the SKA data.

SKA is the world's largest radio telescope soon to be built [1], and will provide a powerful tool for human beings to probe the Universe with unprecedented details and to explore the frontiers of natural sciences [2]. SKA will generate the unparalleled big data flow exceeding anything that astronomers have ever obtained [3,4]. Scientists and engineers across the world are making efforts to handle the data challenges brought by SKA. The SKA Phase 1 (SKA1), about 10% of the total scale, is expected to create data stream flowing into the science data processor at a data rate of 10 tera bit per second [5], which is 100,000 times faster than the average broadband speed of home users. Compared with the SKA pathfinders, e.g., Low Frequency Array (LOFAR) [6], the SKA1 data rate is 2–3 orders of magnitude higher [3]. In order to handle the SKA data flow, astronomers and computer scientists from the International Centre for Radio Astronomy Research (ICRAR, Australia), Shanghai Astronomical Observatory (SHAO, China) and Oak Ridge National Laboratory (ORNL, USA) recently carried out an experiment on the *Summit* supercomputer (being able to achieve 200 peta flops [7]) at the ORNL to simulate the data processing workflow of the SKA1 scale. This is the largest data flow created in the SKA community ever, moreover astronomy data has never been processed at this scale before.

The team first used a simulator OSKAR [8] version 2 to generate datasets of a typical SKA1 observation of 6 hours based on a predicted sky model of the Epoch of Reionization (EoR), and then streamed the data into an ingestion pipeline at a rate of 250 gigabytes per second. That resulted in a visibility data set with size of 2.6 petabyte. The ingested data then went through the imaging pipeline that is composed of a series of processing steps including data averaging, calibration and imaging. The imaging pipeline cost ~3 hours, after that, the pre-calibrated data of 110 terabyte in size were transferred to the Pawsey supercomputer in Perth (Western Australia) and China SKA Regional Centre prototype [9] in Shanghai (China) where the final line cubes were created and presented in a visualisation (see online *Supplementary video*).

The team spent three years on algorithm development, software component integration, code optimization, and verification experiments on some smaller-scale high performance computing (HPC) platforms as well. In October 2019, the workflow was successfully scaled up to run on a maximum of 4560 nodes using 27360 graphic processing units (GPUs). Generating such an enormous amount of visibility data (~2 petabyte) relied on the powerful computing power of the GPUs in the *Summit* system which is probably the only qualified supercomputer. This experiment, at its peak execution, occupied over 99% resource of the entire *Summit* supercomputer, making it one of the largest science applications ever run on the *Summit* system.

The success of the *SKA-Summit* experiment marks an important milestone in the SKA engineering project. For the first time ever, the SKA scale data is processed on a real supercomputer. This experiment illustrates the real data challenge of the SKA: only a single 6-h observation already utilizes almost the whole compute resource of the current fastest supercomputer. On the other hand, the success of the experiment demonstrates that the community has the ability to tackle the SKA data challenge. The team also successfully performed a live demonstration in the 6[th] SKA Engineering Meeting in Shanghai on 2019 November 25, which greatly enhances the confidence of the community of handling the SKA science data processing [10]. This pilot experiment, based on the simulation of the SKA1 low-frequency array configuration, paves the way toward a full implementation of EoR simulation in future, helping astronomers to unveil the mystery of the cosmic evolution. In addition to the present EoR simulation, there are other experiments executed and/or planned to understand the SKA data challenges. The first SKA data challenge has been accomplished in early 2019, with the focus on source finding and classification. Next ones are being prepared to investigate the diverse requirements from different science goals such as HI intensity mapping, transient search, and others.

A lot of innovative designs and developments have been made in order for the workflow deployment to be expanded from tens of nodes to several thousands of nodes. Instead of using the ordinary parallel computing paradigm, an agile workflow management system, *DALiuGE* [11], was designed by the team, and it significantly enhanced the scalability, robustness and flexibility in managing millions of workflow elements. The data-driven or data-triggering manner of *DALiuGE* substantially reduces the idle compute and storage resource, thus significantly increasing the use efficiency. *DALiuGE* can also be applied to other data-intensive scientific workflows, even for non-astronomy research. Another remarkable innovation is the parallel input/output (I/O) framework, ADIOS [12], specially developed for the SKA. The traditional low-level data I/O infrastructures were found to not be able to efficiently manage the massive parallel data reading and writing of the SKA-scale data flow. A considerable amount of effort has been spent on integrating the high performance parallel I/O framework ADIOS into fundamental radio astronomy data processing software packages. The capacity of tera bit per second reading and writing enables ADIOS to be an attractive solution for I/O-bound workflows in the general compute fields.

The SKA-*Summit* experiment shows the importance of multidisciplinary cooperation between astronomy, computer science and others communities. The SKA science cannot be achieved without the joint efforts of talents from multiple fields.

**Conflict of interest**

The authors declare that they have no conflict of interest.

**Acknowledgments**

This work was accomplished on the Summit supercomputer at the Oak Ridge National Laboratory. This work also used resources of China SKA Regional Data Centre prototype (*Nat Astron 2019; 3: 1030*) funded by the National Key R&D Programme of China (2018YFA0404603) and Chinese Academy of Sciences (114231KYSB20170003), the Pawsey Supercomputing Centre funded from the Australian Government and the Government of Western Australia. The simulation tool OSKAR2 is developed by University of Oxford. We thank our colleagues who deeply involved in this work from ICRAR, SHAO, ORNL, and Commonwealth Scientific and Industrial Research Organisation of Australia.

Fig. 1 Illustration of the simulated SKA-Summit workflow

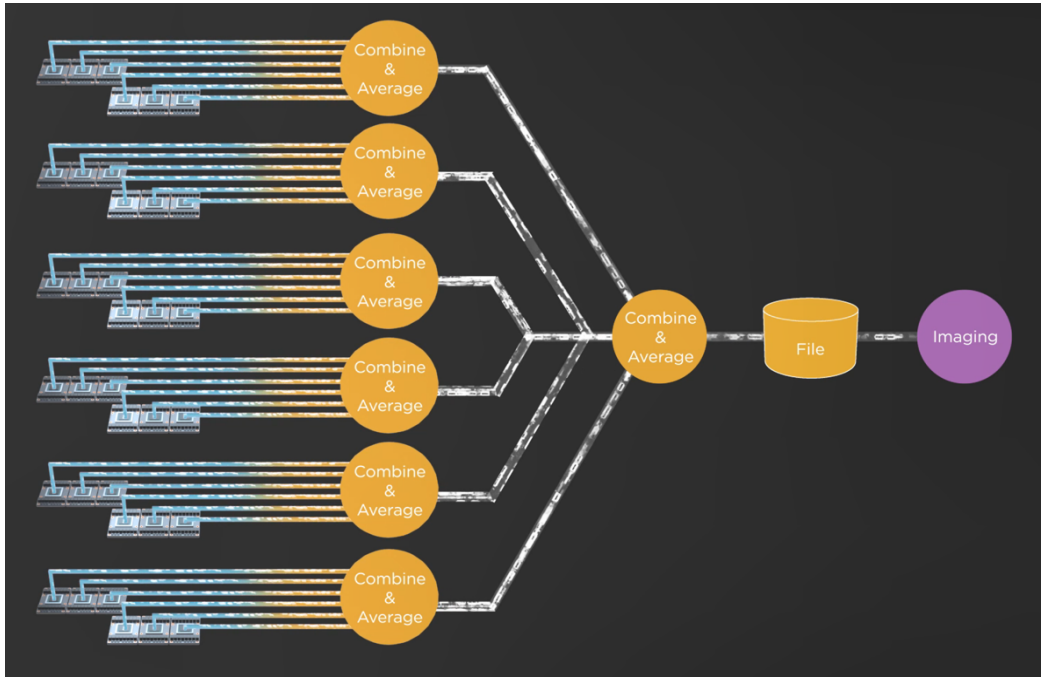

Supplementary Figure: Epoch of Reionization simulation based on SKA1 low array configuration.

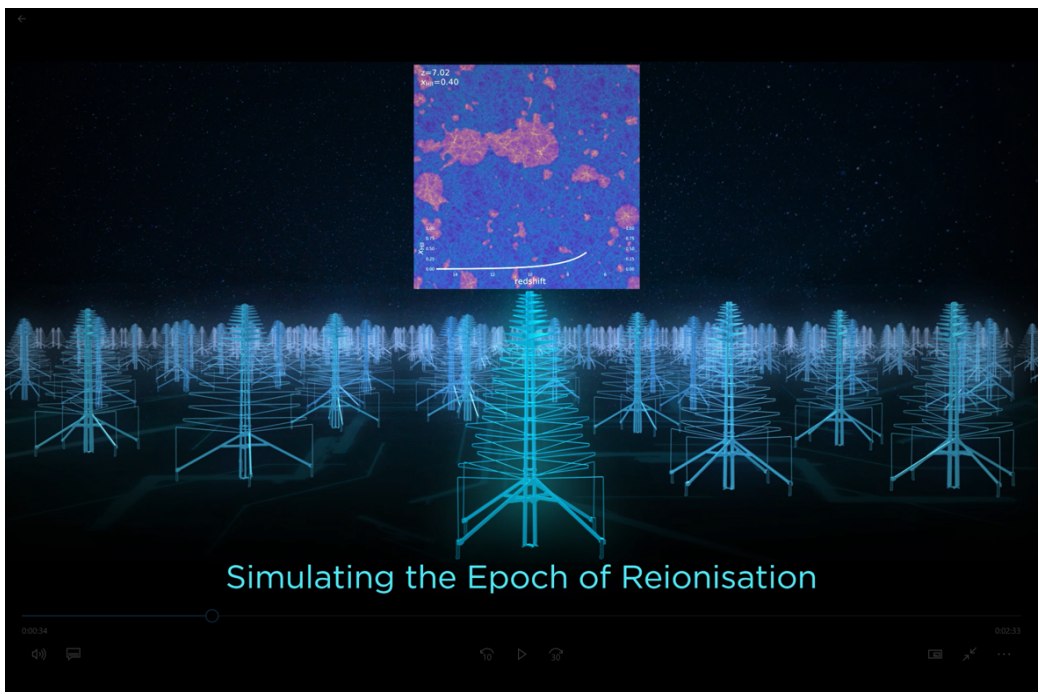

Extras

Supplementary video : Simulating SKA1-Low processing using the world's biggest supercomputer
 https://www.sciencedirect.com/science/article/pii/S2095927319307157?via%3Dihub

Vitae

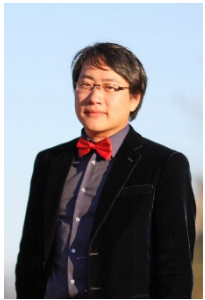

Ruonan Wang is a Software Engineer in the Scientific Data Group, Computer Science and Mathematics Division at the Oak Ridge National Laboratory of the USA. He obtained a Ph.D. degree at the University of Western Australia, and then worked at the International Centre for Radio Astronomy Research (ICRAR). His work focuses on data management systems and data I/O middleware design for the future world's largest radio telescope, Square Kilometer Array (SKA).

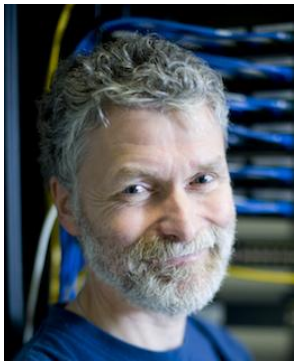

Andreas Wicenec is a professor at the University of Western Australia, leading the Data Intensive Astronomy Program of the International Centre for Radio Astronomy Research (ICRAR) to research, design and implement Petabyte scale data flows and high performance scientific computing for the Murchison Wide Field Array (MWA), the Australian SKA Pathfinder (ASKAP) and the Square Kilometer Array (SKA).

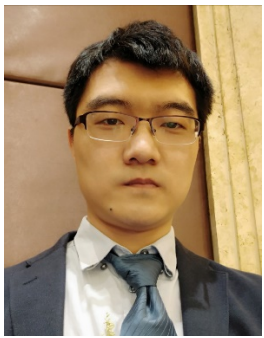

Tao An is a professor at the Shanghai Astronomical Observatory of the Chinese Academy of Sciences. His research fields are astrophysics, radio astronomy, and very long baseline interferometry (VLBI). He is Member of Square Kilometre Array (SKA) Regional Centre Streering Committee and International Astronomical Society (IAU) Commission B4, and co-chair of SKA VLBI science working group. He is leading the China SKA Regional Centre prototype construction.